\documentclass{amsart}
\usepackage{a4wide,times}

\usepackage{listings}
\newcommand{\CODE}[2]{\lstinputlisting[firstline=#2]{#1}}
\newcommand{\CODElast}[3]{\lstinputlisting[firstline=#2,lastline=#3]{#1}}

\begin{document}

\title[Methods of Matrix Multiplication]{Methods of Matrix Multiplication\\[1ex]\normalfont An Overview of Several Methods and their Implementation}
\author[Ivo Hedtke]{Ivo Hedtke\\Version 1 (\today{})}
\address{Ivo Hedtke, Research Group: Data Structures and Efficient Algorithms, Institute for Computer Science, Martin-Luther-Universit\"at Halle-Wittenberg, Von-Seckendorff-Platz 1, 06120 Halle, Germany}
\email{Ivo.Hedtke@uni-jena.de}

\begin{abstract}
In this overview article we present several methods for multiplying matrices and the implementation of these methods in C. Also a little test program is given to compare their running time and the numerical stability.

The methods are: naive method, naive method working on arrays, naive method with the \textsc{Kahan} trick, three methods with loop unrolling, winograd method and the scaled variant, original \textsc{Strassen} method and the \textsc{Strassen}-\textsc{Winograd} variant.
\end{abstract}
\maketitle

\textbf{Please note, that this is the FIRST version. The algorithms are not well tested and the implementation is not optimized. If you like to join the project, please contact me.}

\textbf{The collection of algorithms and this document will be updated from time to time.}

\tableofcontents

In the following text, $A$ denotes a $N\times P$ matrix and $B$ denotes a $P\times M$ matrix with entries of type \texttt{double}. The implementation is written in C based on the standard C99. The code and the current version of this document can be found here:

\texttt{http://www2.informatik.uni-halle.de/da/hedtke/overview-momm/}.

Feel free to use the code for anything you want. DO NOT USE COMPILER OPTIMIZATION!

\newpage\section{Auxiliary Routines}
\subsection{Plus} According to the definition of the $+$ operator for matrices
\[
(A+B)_{ij} = A_{ij} + B_{ij}
\]
we use a simple straightforward method to compute $A+B$.
\CODE{Plus.c}{3}

\subsection{Minus} Like above we use a straightforward method to compute
\[
(A-B)_{ij}= A_{ij} - B_{ij}.
\]
\CODE{Minus.c}{3}

\subsection{max} As an often used auxiliary function we implement
\[
\max(x,y) = \begin{cases}
x & x \geq y\\
y & \text{otherwise}
\end{cases}
\]
for \texttt{int}eger and \texttt{double} input arguments.
\CODE{max.c}{3}

\subsection{NormInf} In the scaled variant of \textsc{Winograd}s algorithm we need the $\infty$-norm
\[
\|A\|_\infty = \max_{i=1,\ldots,N}\left\{\sum_{j=1}^P |A_{ij}|\right\}
\]
of a matrix $A$. We also use it to compare the numerical stability of the several methods in the tests.
\CODE{NormInf.c}{3}

\subsection{MultiplyWithScalar} Also in the scaled variant of \textsc{Winograd}s algorithm we need the product of a matrix $A$ and a scalar $\alpha$
\[
(\alpha A)_{ij} = \alpha A_{ij}.
\]
\CODE{MultiplyWithScalar.c}{3}

\newpage\section{Methods of Matrix Multiplication}

\subsection{The naive method} The naive method of matrix multiplication is given by the definiton itself:
\[(AB)_{ij}=\sum_{k=1}^P A_{ik}B_{k j}.\]

\subsubsection{NaivStandard} First we implement this straightforward but we work with an auxiliary variable \texttt{aux} for the sum.
\CODE{NaivStandard.c}{3}

\subsubsection{NaivOnArray} A common error is to work all the time on the arrays itself instead of using an auxiliary variable.  We only present this method to include it in the tests.
\CODE{NaivOnArray.c}{3}

\subsubsection{NaivKahan} To improve the numerical stability of the process, we use the \textsc{Kahan} trick (see \cite{kahan}) for the summation.
Instead of computing a sum like
\[\sum_{i=1}^n x_i\]
with
\begin{verbatim}
double sum = 0.0;
for( int i = 1; i <= n; i++){
  sum += x[i];
}
\end{verbatim}
we use
\begin{verbatim}
double t;
double sum = 0.0;
double err = 0.0;
for( int i = 1; i <= n; i++){
  err = err + x[i];
  t = sum + err;
  err = (sum - t) + err;
  sum = t;
}
\end{verbatim}

\CODE{NaivKahan.c}{8}

\subsection{Loop unrolling}

The method of \emph{loop unrolling} doesn't reduce the abstract complexity of the matrix mutliplication problem. It tries to reduce the overhead from the \texttt{for} loops and speed up the real computational work.

\subsubsection{NaivLoopUnrollingTwo} Instead of computing
\[
(AB)_{ij}=\sum_{k=1}^P A_{ik}B_{kj},
\]
we use
\begin{align*}
(AB)_{ij} &=\sum_{k=1}^{P/2} A_{i,2k-1}B_{2k-1,j} + A_{i,2k}B_{2k,j}, \qquad \text{or}\\
(AB)_{ij} & =\left(\sum_{k=1}^{\lfloor P/2 \rfloor} A_{i,2k-1}B_{2k-1,j} + A_{i,2k}B_{2k,j}\right)
+ A_{i,P}B_{P,j},
\end{align*}
if $P$ is even or odd, resp.

\CODE{NaivLoopUnrollingTwo.c}{3}

\subsubsection{NaivLoopUnrollingThree} Like above we compute
\[
(AB)_{ij} =\sum_{k=1}^{P/3} A_{i,3k-2}B_{3k-2,j} +  A_{i,3k-1}B_{3k-1,j} + A_{i,2k}B_{2k,j}
\]
instead of
\[
(AB)_{ij}=\sum_{k=1}^P A_{ik}B_{kj},
\]
if $P$ is divisible by $3$ (otherwise we use correction terms like above).

\CODE{NaivLoopUnrollingThree.c}{3}

\subsubsection{NaivLoopUnrollingFour}~

\CODE{NaivLoopUnrollingFour.c}{3}

\subsection{Winograd's algorithm}
\subsubsection{WinogradOriginal} From \textsc{Shmuel Winograd} we know a method that precomputes some values and reuse it in the computation of the entries of the result matrix. It is based on the fact that
\[
(AB)_{ik}=\sum_{j=1}^{P/2} (A_{i,2j-1} + B_{2j,k})(A_{i,2j} + B_{2j-1,k}) - \underbrace{\sum_{j=1}^{P/2}A_{i,2j-1}A_{i,2j}}_{=:y_i} -
\underbrace{\sum_{j=1}^{P/2}B_{2j-1,k}B_{2j,k}}_{=:z_k}
\]
if $P$ is even. In the case that $P$ is odd, we use $\lfloor P/2\rfloor$ instead of $P/2$ and add the missing product $A_{i,P}B_{B,k}$ to each entry of the result matrix.

In the implementation we use \texttt{upsilon} to indicate if $P$ is even and \texttt{gamma} as $\lfloor P/2 \rfloor$.
\CODE{WinogradOriginal.c}{3}

\subsubsection{WinogradScaled} From \textsc{R. P. Brent} we know (see \cite{brent}) that the error of \textsc{Winograd}'s algorithm is
\[
\|E\|\leq 2^{-\tau}\frac{N^2+12N-8}{4}(\|A\|+\|B\|)^2,
\]
where $\tau$ is a parameter of the underlying number modell (defined by \textsc{Wilkinson} in \cite{wilk}) and $N$ is size of the matrices (in the case that $N=P=M$). 
Now suppose that $\|A\|/\|B\| = k$, then $(\|A\|+\|B\|)^2 = (k+2+1/k)\|A\|\|B\|$, and so
\[
\|E\|\leq 2^{-\tau}\frac{N^2+12N-8}{4}(k+2+1/k)\|A\|\|B\|.
\]
According to \textsc{Brent} it is always possible to find an integer $\lambda$ such that
\[
1/2 \leq \frac{2^\lambda \|A\|}{2^{-\lambda}\|B\|} \leq 2.
\]
And because of
\[\max_{\frac12\leq k \leq 2} k+2+1/k=9/2\]
he gets the bound
\[
\|E\|\leq 2^{-\tau}\cdot \frac98 \cdot (N^2+12N-8)\cdot \|A\|\cdot\|B\|
\]
when multiplying the matrices $2^\lambda A$ and $2^{-\lambda}B$ (which doesn't change the result: $(2^\lambda A) \cdot (2^{-\lambda}B) = 2^\lambda 2^{-\lambda} A B=AB$).
\CODE{WinogradScaled.c}{7}

\subsection{Strassen's algorithm}

The matrix multiplication algorithm from \textsc{Volker Strassen} is very famous. We present two variants.

\subsubsection{StrassenNaiv} In his paper \emph{Gaussian Elimination is not Optimal} (see \cite{strassen}) \textsc{Strassen} developed a recursive algorithm $\alpha_{m,k}$ for matrix multiplication for square matrices of order $m2^k$, where $k,m\in\mathbb N$. Let $\alpha_{m,0}$ be the naive algorithm for matrix multiplication. We assume that $\alpha_{m,k}$ is known, then we define $\alpha_{m,k+1}$ as follows:

If $A$, $B$ are matrices of order $m2^{k+1}$ to be multiplied, we write
\begin{gather}\label{eq:StrassenZerlegen}
A=\begin{bmatrix}
A_{11} & A_{12}\\
A_{21} & A_{22}
\end{bmatrix},\quad 
B=\begin{bmatrix}
B_{11} & B_{12}\\
B_{21} & B_{22}
\end{bmatrix}\quad
\text{and}\quad
C:=AB=\begin{bmatrix}
C_{11} & C_{12}\\
C_{21} & C_{22}
\end{bmatrix},
\end{gather}
where $A_{ij}$, $B_{ij}$ and $C_{ij}$ are matrices of order $m2^k$. With the following auxiliary matrices
\begin{xalignat*}{2}
H_1 &:= (A_{11} + A_{22})(B_{11} + B_{22}) &
H_2 &:=(A_{21} + A_{22})B_{11} \\
H_3 &:=A_{11}(B_{12} - B_{22}) &
H_4 &:= A_{22}(B_{21} - B_{11}) \\
H_5 &:=(A_{11} + A_{12})B_{22} & 
H_6 &:=(A_{21} - A_{11})(B_{11} + B_{12})\\
H_7 &:= (A_{12} - A_{22})(B_{21} + B_{22})
\intertext{of order $m2^k$ computed using $\alpha_{m,k}$ for matrix multiplication and the usual algorithm for addition and subtraction of matrices we get}
C_{11} & = H_1 + H_4 - H_5 + H_7 &
C_{12} & = H_3 + H_5\\
C_{21} & = H_2 + H_4 &
C_{22} & = H_1 + H_3 - H_2 + H_6.
\end{xalignat*}
as the entries of the result matrix $C$.

To use the algorithm above we define $X:=\max\{N,P,M\}$ and set $k:=\lfloor\log_2 X\rfloor -4$ and $m:=\lfloor X2^{-k}\rfloor + 1$. Now with $Y:=m2^k$ we define matrices $\tilde A$ and $\tilde B$ of size $Y\times Y$. The main idea is, to embed the given  matrices into $Y \times Y$ matrices, by adding zero rows and columns. Now we use the \textsc{Strassen} algorithm to compute $\tilde A\tilde B$ and extract the result $AB$ from it, by removing the additional rows and colums.

In the implementation we use \texttt{ResultPart}=$C$, \texttt{Aux1}=$H_1$, \ldots, \texttt{Aux7}=$H_7$, \texttt{AuxResult}=$\tilde A \tilde B$,\linebreak\texttt{MaxSize}=$X$ and \texttt{NewSize}=$Y$.

\CODE{StrassenNaiv.c}{219}
\CODElast{StrassenNaiv.c}{3}{214}

\subsubsection{StrassenWinograd} In the paper \emph{Efficient Procedures for using Matrix Algorithms} (see \cite{fischer}) \textsc{Patrick C. Fischer} and \textsc{Robert L. Probert} discussed an idea of \textsc{Shmuel Winograd} which reduces the number of used additions to $15$ (instead of $18$ in \emph{StrassenNaiv}).

Like above, we define a recursive algorithm $\alpha$ for multiplication of square matrices of order $m2^{k+1}$.
Let $A$ and $B$ be matrices of this size. We assume that $\alpha$ is already known for the order $m2^k$.
We define $C:=AB$ and decompose $A$, $B$ and $C$ according to Equation \eqref{eq:StrassenZerlegen}. 
We define\begin{align*}
A_1&:=A_{11}-A_{21} & B_1&:=B_{22} - B_{12}\\
A_2&:=A_{22} - A_1  & B_2&:=B_1 + B_{11}
\intertext{and}
H_1 &=A_{11}B_{11} & H_5 &= A_1B_1\\
H_2 &=A_{12}B_{21} & H_6 &= (A_{12} - A_{2})B_{22}\\
H_3 &=A_2B_2 & H_7&=A_{22}(B_{21} - B_{2})\\
H_4 &=(A_{21} + A_{22})(B_{12} - B_{11}).
\intertext{With}
H_8&:=H_1+H_3 & H_9 &:= H_8 + H_4
\intertext{we finally get}
C_{11} &= H_1 + H_2 & C_{12} &= H_9 + H_6\\
C_{21}&= H_8 + H_5 + H_7 & C_{22} &= H_9 + H_5.
\end{align*}

\CODE{StrassenWinograd.c}{229}
\CODElast{StrassenWinograd.c}{3}{225}

\newpage\section{Tests}

To test the algorithms we compute the product of a random square matrix $A$ with its scaled variant $B:=8A$. The test routine is part of the package of C files. Is has the command line options
\begin{itemize}
\item \texttt{-O} matrix size, \emph{standard:} 400
\item \texttt{-R} number of repeats, \emph{standard:} 10
\end{itemize}

\begin{verbatim}
+-----------------------------------------------------------------------+
|            TIME TEST FOR METHODS OF MATRIX MULTIPLICATION             |
|                                                                       |
| C = A*(8A), where A is a (n x n) random matrix with n =        200    |
+-----------------------------------------------------------------------+
| method                      |       time (sec) |       NormInf( N-C ) |
+-----------------------------------------------------------------------+
| N := NaivKahan              |         0.146027 |                      |
| NaivStandard                |         0.083263 |         0.0000000000 |
| NaivOnArray                 |         0.121735 |         0.0000000000 |
| NaivLoopUnrollingTwo        |         0.072981 |         0.0000000000 |
| NaivLoopUnrollingThree      |         0.068356 |         0.0000000000 |
| NaivLoopUnrollingFour       |         0.067080 |         0.0000000000 |
| StrassenNaiv                |         0.077584 |         0.0000000001 |
| StrassenWinograd            |         0.080180 |         0.0000000000 |
| WinogradOriginal            |         0.073193 |         0.0000000001 |
| WinogradScaled              |         0.061350 |         0.0000000001 |
+-----------------------------------------------------------------------+

+-----------------------------------------------------------------------+
|            TIME TEST FOR METHODS OF MATRIX MULTIPLICATION             |
|                                                                       |
| C = A*(8A), where A is a (n x n) random matrix with n =        800    |
+-----------------------------------------------------------------------+
| method                      |       time (sec) |       NormInf( N-C ) |
+-----------------------------------------------------------------------+
| N := NaivKahan              |        11.817480 |                      |
| NaivStandard                |         7.370365 |         0.0000000009 |
| NaivOnArray                 |        10.828352 |         0.0000000009 |
| NaivLoopUnrollingTwo        |         6.914036 |         0.0000000006 |
| NaivLoopUnrollingThree      |         6.605404 |         0.0000000005 |
| NaivLoopUnrollingFour       |         6.480659 |         0.0000000004 |
| StrassenNaiv                |         3.806864 |         0.0000000022 |
| StrassenWinograd            |         3.978543 |         0.0000000010 |
| WinogradOriginal            |         6.892913 |         0.0000000036 |
| WinogradScaled              |         5.781587 |         0.0000000022 |
+-----------------------------------------------------------------------+
\end{verbatim}

\end{document}